\documentclass[12pt]{iopart}
\usepackage{iopams} 
\usepackage{wasysym}
\usepackage{epsfig}
\def\3{\ss}
\def\3{\ss} 
\def\Journal#1#2#3#4{{#1} {\bf #2}, #3 (#4)}

\def\PRD{{\em Phys. Rev.} D}

\def\EPC{{\em Eur. Phys. J.} C}
\newcommand{\csm}{Colour Singlet Model}
\newcommand{\dsdz}{$d\sigma/dz$}
\newcommand{\ccbar}{c\overline{c}}
\newcommand{\ppbar}{$p\overline{p}$}

\newcommand{\ra}{\rightarrow}
\newcommand{\picb}{\mbox{pb}^{-1}}
\newcommand{\gev}{\,\mbox{\rm GeV}}
\newcommand{\GeV}{\,\mbox{\rm GeV}}

\newcommand{\gevt}{\,\mbox{\GeV$^2$}}

\newcommand{\qsq}{\ensuremath{Q^2} }
\newcommand{\ptt}{\ensuremath{p_t^2} }
\newcommand{\ptr}{\ensuremath{p_t} }
\newcommand{\ptpsi}{\ensuremath{p_{t,\psi}}}
\newcommand{\wgp}{\ensuremath{W_{\gamma p}}}
\newcommand{\jpsi}{$J/\psi$}
\begin{document}

\title[Production of $J/\psi$ Mesons at HERA]{Production of $J/\psi$ Mesons at HERA}

\author{Beate Naroska\\on behalf of the H1 and ZEUS collaborations}  

\address{Institut f\"ur Experimentalphysik\\
Luruper Chaussee 149\\D 22761 Hamburg\\
E-mail:naroska@mail.desy.de}

\begin{abstract}
Inelastic and diffractive production of $J/\psi$ 
mesons at HERA is reviewed.
The data on inelastic photoproduction are described well within errors
by the Colour Singlet Model in next-to-leading order. A search for
colour octet processes predicted within the NRQCD/factorisation
approach is conducted in many regions of phase space. No unambiguous
evidence has been found to date. Diffractive elastic production of
$J/\psi$ mesons has been measured in the limit of photoproduction
($Q^2 \simeq 0)$ up to the highest photon proton center of mass energies. 
The increase of the
cross section is described by pQCD models. At larger $Q^2,\ 2 \leq Q^2 \leq
100\ \mbox{GeV}^2$, the $W_{\gamma p}$ dependence is found to be similar
to that observed in photoproduction.
First analyses of data at high $t$, $|t| \leq 20\ \mbox{GeV}^2,$ yield a
powerlike dependence on $|t|$. A LO BFKL calculation gives a good
description of the data.

\end{abstract}


\maketitle

\section{Introduction}
The main interest in leptoproduction of \jpsi\ mesons is a clarification of 
the production process. In the inelastic regime, many models have been proposed, and
predictions  within the theoretically favoured NRQCD/factorisation approach 
could not yet be identified nor could they be ruled out. 
In the diffractive regime, models based on pQCD have proven to work in 
principle and  more detailed studies are underway experimentally and 
theoretically.

This review will cover $J/\psi$ production at HERA in inelastic and 
diffractive processes.
Inelastic $J/\psi$ production at HERA has been studied in the limit of 
photoproduction ($Q^2 \rightarrow 0)$ and for $Q^2 \gtrsim$ 2 GeV$^2$. 
Data at high $Q^2$ were published by the H1 collaboration~\cite{Adloff:1999zs}
and new theoretical calculations are discussed in L. Zwirner's contribution~\cite{zwirner}.
Here I  concentrate on photoproduction.  

Diffractive elastic $J/\psi$ production is studied with large statistics 
in the limit of photoproduction
and also at $Q^2 \gtrsim$ 2 GeV$^2$, where the \wgp\ and $Q^2$ dependences are 
analysed. Proton dissociative data at high values of $t$, 
the momentum transfer squared at the proton vertex, are also analysed.

\begin{figure}[b!]
\unitlength1.0cm
\begin{picture}(20,3.5)
\put(0.,0.){\epsfig{file=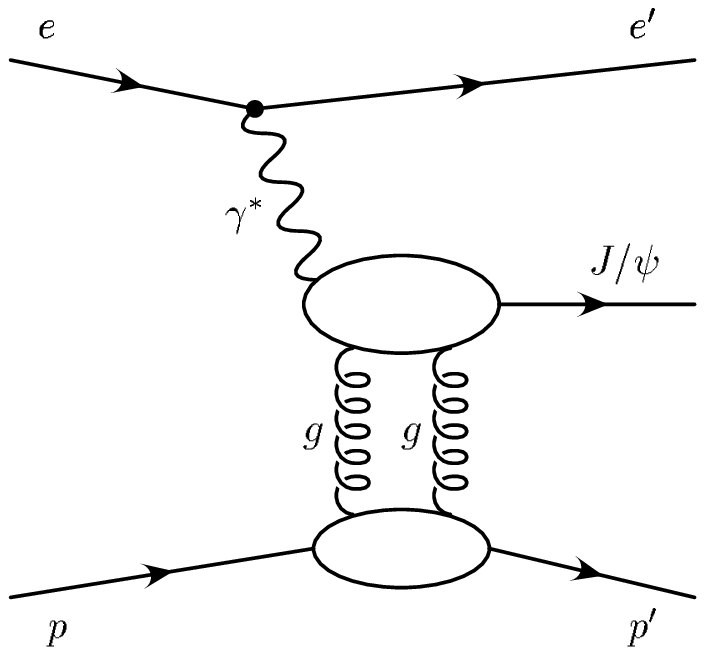,width=4.5cm,clip=}}
\put(2.,0.){(a)}
\put(4.5,-0.5){\epsfig{file=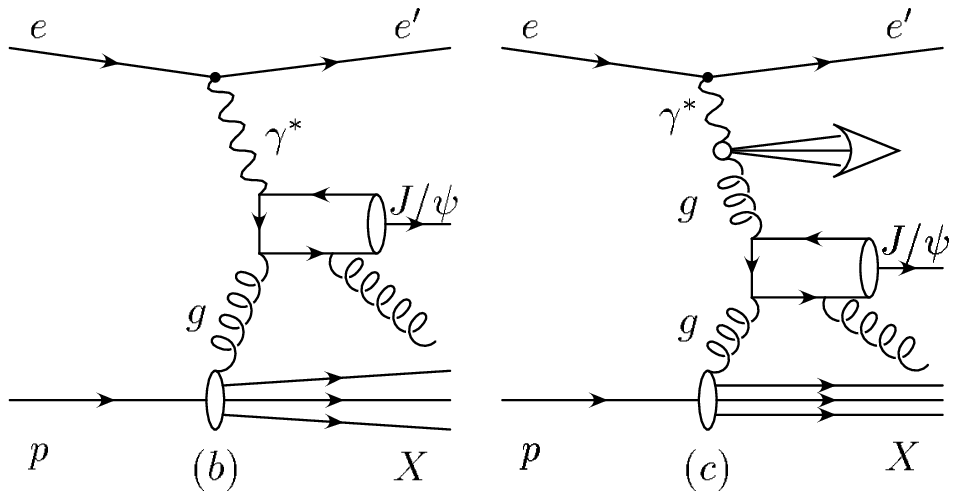,width=9cm}}
\end{picture} 
\caption{\label{fig1}
 Generic diagrams for $J/\psi$ production mechanisms at HERA: a) diffractive elastic process 
with two gluon exchange. Inelastic processes via boson gluon fusion 
with b) direct and c) resolved photons.}
\end{figure}

\section{Inelastic Production of \jpsi~Mesons}
The data from the H1\cite{h1prel} and
ZEUS collaborations~\cite{zeusprel} on inelastic \jpsi\ photoproduction 
are based on integrated luminosities of $22\,\picb$ and $38\,\picb$,
respectively, collected in the years 1996/97. The kinematic regions are
characterized by $Q^2 \leq 1\ \mbox{GeV}^2, \ 50 < 
W_{\gamma p} < 180\ \mbox{GeV},\ 0.4 < z < 0.9$ for the 
ZEUS data and $60 < W_{\gamma p} < 180\
\mbox{GeV}^2,\ 0.3 < z < 0.9$ for H1. The usual kinematic quantities
 $Q^2 = -(k - k')^2$ and  $W^2_{\gamma p} = (q + P)^2$ are used, where
 $k,\ k',\ q\ \mbox{and}\ P$ denote the four-momenta of the incoming
and outgoing electron, of the exchanged photon and the incoming proton, 
respectively. An important variable is

$$z = \frac{q\cdot P_\psi}{q\cdot P}$$
\noindent
where $P_\psi$ is the four-momentum of the $J/\psi$ meson. In the proton 
rest frame $z$ is the relative energy
of the $J/\psi$ with respect to the photon, $z\approx \frac{E_\psi^*}{E_\gamma^*}$. The variable $z$ is used to distinguish
diffractive processes, where $z \approx 1$, from inelastic processes, where
$z \lesssim 0.9$. 

The inelastic production
process at HERA is dominated by boson gluon fusion, see Figure~\ref{fig1}b,c. 
At medium $z$, direct processes dominate, where the photon couples directly 
to the charm quark, 
while at small $z$ resolved photon processes are expected to contribute, 
where the photon interacts via its hadronic component.

\begin{figure}[b!]
\unitlength1.0cm
\begin{picture}(10,9.5) 
\unitlength1.0cm
\put(-0.5,0.){\epsfig{file=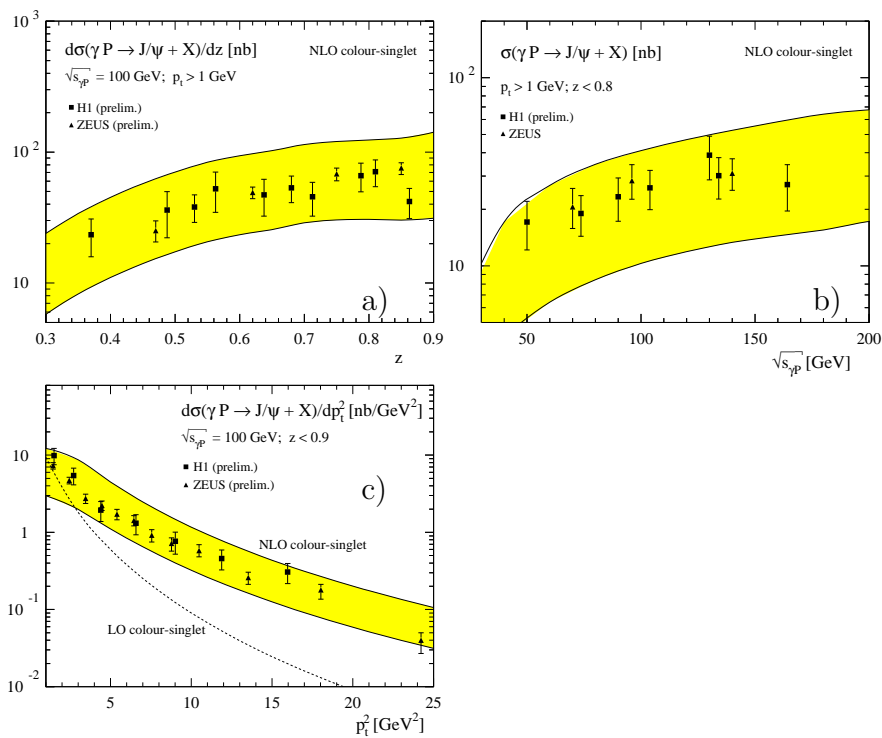,width=12.cm,clip=}}
\put(4.5,6.){a)}
\put(10.5,6.){b)}
\put(4.5,3.5){c)}
\end{picture}
\vspace{-0.7cm}
\caption{\label{fig2}
Cross sections as functions of $z$, \protect\wgp, and \protect\ptpsi$^2$. 
The data from H1~\cite{h1prel} and ZEUS~\cite{z94,zeusprel} are shown with NLO CSM calculations. The band 
reflects the uncertainty due to the 
charm mass ($1.3<m_c<1.5\,$GeV) and $0.1175<\alpha_s(M_Z)<0.1225$. The renormalisation and factorisation scales are $\mu^2=2\,m_c^2$. (Figures from \cite{Kraemer}.)}
\end{figure}

The theoretical description of \jpsi\ meson production has undergone a change 
in the last few years. The Colour Singlet Model (CSM) \cite{csm} has been replaced by an
effective field theory approach: non--relativistic QCD (NRQCD) and
factorisation \cite{bbl95}. This approach contains the CSM 
but predicts additional contributions. 
In the CSM, the $c\overline{c}$ pair is  produced in the hard scattering 
 with the observed
quantum numbers of the $J/\psi$ meson. The colour singlet state is achieved 
by emission of an additional gluon, 
i.e. the process is $\gamma g\ra g+\ccbar\left[1,^3S_1\right]$\footnote{The `1' or `8' denote the colour state of the  $c\overline{c}$ pair and spectroscopic notation is used for the angular momentum, $^{2S+1}L_J$.}.
In the NRQCD/factorisation approach the $c\overline{c}$ pair can also be in a 
colour octet state where different angular momentum states are also  allowed.

At HERA we are in a unique position since for photoproduction 
full next-to-leading order calculations
are available in the colour singlet model, which reduces 
the uncertainties in the normalisation of the predictions. So we will first 
show a comparison of the
data with these calculations before we address possible
contributions of colour octet (CO) intermediate states.

\begin{figure}[b!]
\unitlength1.0cm
\begin{picture}(20,10) 
\put(0,0){\epsfig{file=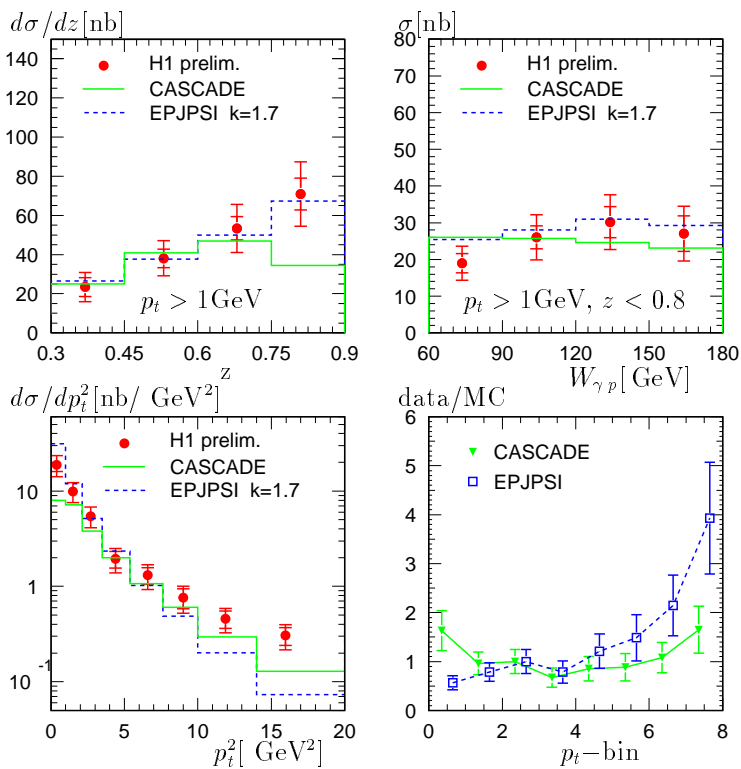,width=10cm,clip=}}
\end{picture} 
\caption{\label{fig5} 
Differential cross sections for photoproduction of \jpsi\ mesons as functions 
of $z$, \wgp\ and \ptt\ and the 
ratio of Data/MC for the \ptt\ bins, normalized in the third bin. The predictions are from the CCFM Monte 
Carlo generator CASCADE and the Colour Singlet Model EPJPSI (the latter scaled by a factor 1.7) are compared to  H1 data with \ptr$>1$\gev and $z<0.8$.}
\end{figure}

\subsection{Comparison of data to Colour Singlet calculations}

In Figure \ref{fig2}a,b the $z$ and the $W_{\gamma p}$
distribution are shown for data from ZEUS~\cite{zeusprel} and H1~\cite{h1prel} 
which agree well with
each other. The prediction of the CSM in next-to-leading order~\cite{Kraemer} is overlaid. The
band reflects the major theoretical uncertainties as detailed in the figure 
caption. A good description of the data is achieved. The necessity of including
NLO contributions is demonstrated in the distribution of $p_{t,\psi}^2$
in Figure \ref{fig2}c. The theoretical band including NLO contributions 
describes the data well and is,
at a $p_{t,\psi} \sim 4\ \mbox{GeV}$, roughly a factor of 10 above the LO 
prediction.
The discrepancy found in  $p\overline{p}\rightarrow J/\psi + X$
at values of $p_t \sim 20 - 30\ \mbox{GeV}$ between the data and the
 LO CSM calculation which led to the introduction of CO
contributions within the NRQCD/factorisation approach was
approximately a factor $\sim$ 40~\cite{fail}. 

A different approach to describe the data within the CSM is shown in 
Figure \ref{fig5}. Here the H1 data are shown with the results of two Monte 
Carlo implementations of the \csm\ in leading order. 
The dashed histograms represent the prediction from  the EPJPSI~\cite{epjpsi}
program which uses ``standard'' parton distributions integrated over the 
transverse momentum and evolved with DGLAP equations.
The program CASCADE \cite{cascade} uses the CCFM evolution equation 
and ``unintegrated'' parton distributions. These have a distribution in 
transverse momentum, $k_t$, and are obtained from a fit to 
H1 data on $F_2$.
The \jpsi\ data in Figure \ref{fig5}(note that now a cut $z < 0.8$ has been 
applied) are in general well 
described by the CASCADE model except at very large $z$ values, where relativistic corrections are missing. The $\ptpsi^2$
distribution of CASCADE agrees better with the data than EPJPSI, but
there is still a systematic discrepancy towards high $p_t$. 

\begin{figure}[b!]
\unitlength1.0cm
\begin{picture}(10,7.5) 
\put(-1,0){\epsfig{file=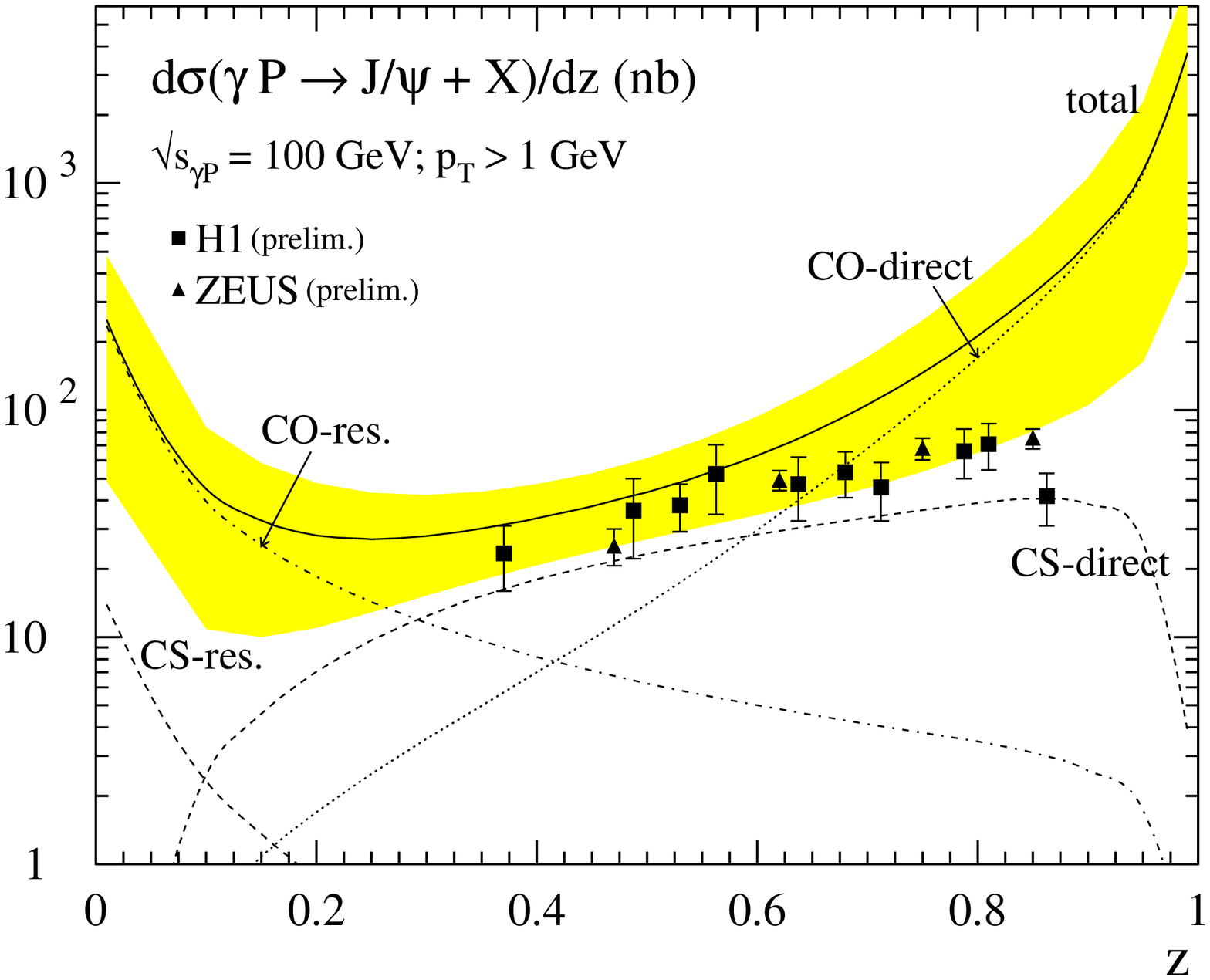,width=0.6\textwidth}}
\put(4.,0.){a)}
\put(7.5,0){\epsfig{file=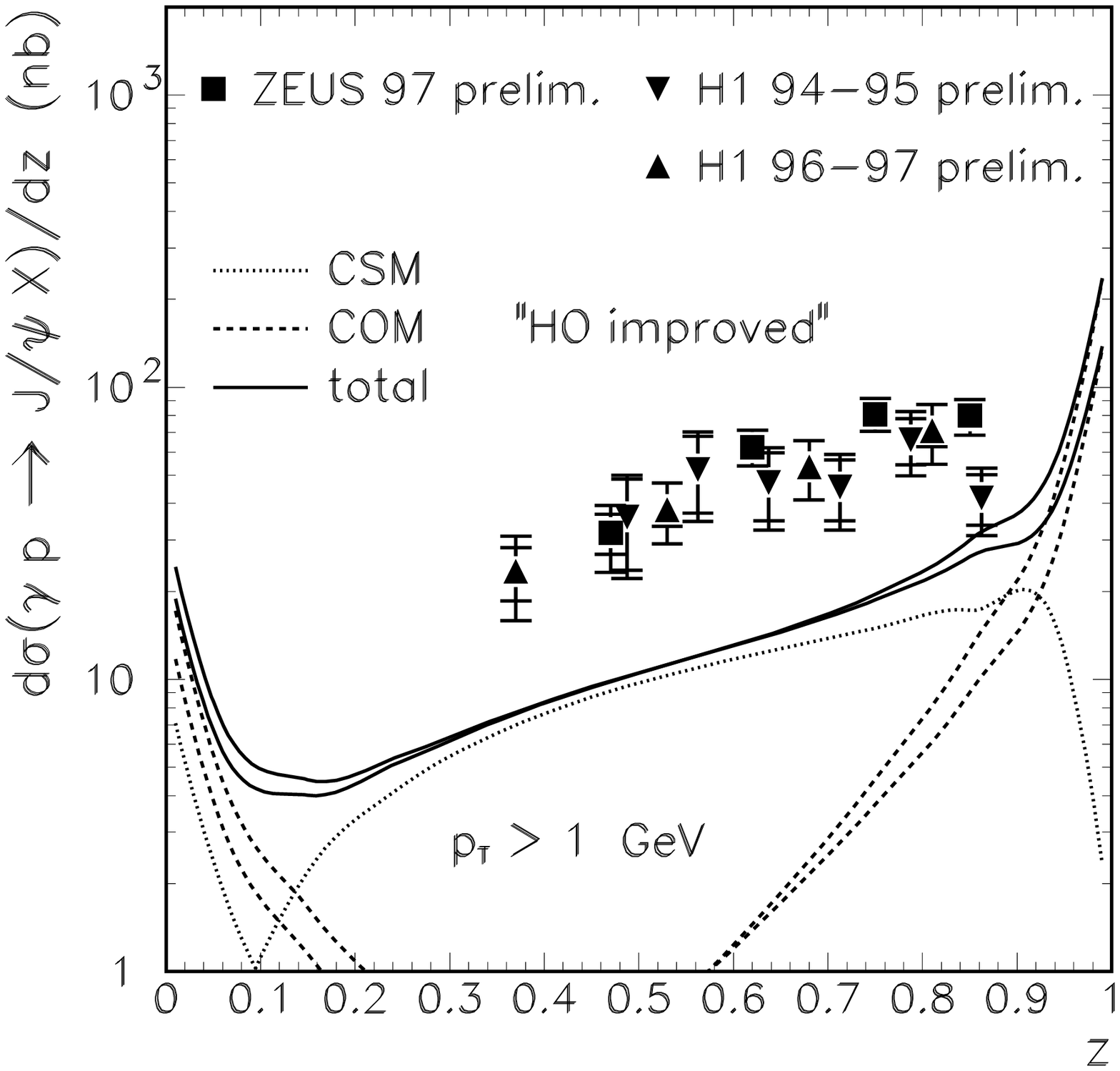,width=0.5\textwidth,
bbllx=0pt,bblly=20pt,bburx=500pt,bbury=500pt,clip=}}
\put(10.,0.){b)}
\end{picture}
 \caption{\label{fig4}
 Differential cross section for photoproduction of \jpsi\ mesons as a function 
of $z$. In a) the predictions from Kraemer~\cite{Kraemer} are 
shown, in b) those from Kniehl et al.~\cite{kniehl}. 
Both predictions include direct 
and resolved contributions, but use different parameters. 
The band in a) depicts the uncertainties which are mainly due to the long range matrix elements. In b) the uncertainties are indicated by the upper and lower curves.}
\end{figure}

\begin{figure}
\unitlength1.0cm
\begin{picture}(20,8.5)
\put(0.,-.5){\epsfig{file=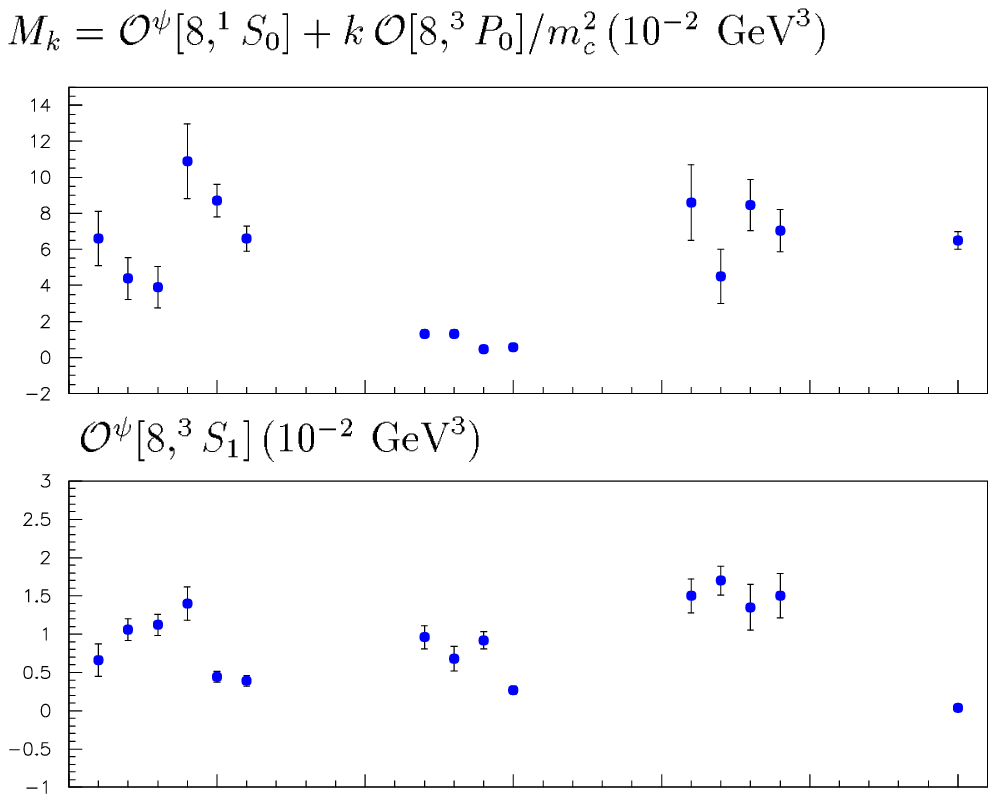,width=12cm,clip=}}
\end{picture}
 \caption{\label{me}
Long range matrix elements as compiled in \cite{Kraemer}. 
The values are extracted from the transverse momentum distribution in 
\ppbar\ \jpsi\ production using different theoretical approximations. 
Only statistical errors are shown. For more details see text. }
\end{figure}

\subsection{Search for Colour Octet Contributions}

In Figure \ref{fig4}a the same differential cross sections, 
$d\sigma/dz$ from the H1 and ZEUS data as before, are shown again as function 
of $z$. The theoretical predictions within 
the NRQCD/factorisation approach in leading order~\cite{Kraemer} are shown 
in comparison. 
These include the direct photon contributions as well as resolved processes.
At $z \gtrsim 0.3$, direct photon contributions dominate and the
CO contribution is dominant at $z \gtrsim$ 0.6. 
The uncertainties of the summed prediction, shown as a band, 
are mainly due to the uncertainties of the long range matrix elements, the
parameters which, in the NRQCD/factorisation approach describe the transition 
of the $\ccbar$\ state to a physical \jpsi\ meson. 

 The long range matrix elements are denoted as e.g. 
${\cal O}^{\psi}\left[8,^3S_1\right]$ to describe the transition of the 
$\ccbar$ state in a colour octet configuration with angular momentum state 
$^3S_1$. They are not calculable but are believed to be universal.
Numerical values have been determined in a number of theoretical 
analyses of $p\overline{p}$ data from the CDF collaboration using different 
approximations (see \cite{Kraemer} for a summary and references).
The most important matrix elements for photoproduction are 
${\cal O}^{\psi}[8, ^1S_0]$ and $\left[8,^3P_J\right]$ which are derived 
in the form of a linear combination $M_k={\cal O}^{\psi}[8,^1S_0]+k\,{\cal O}^{\psi}[8,^3P_0]/m_c^2$, where $k$ is a parameter of order 3.
In Figure \ref{me},  the values for $M_k$ (upper part) and the matrix element 
${\cal O}^{\psi}[8,^3S_1]$ (lower part) extracted by a few groups using 
different approximations are shown following a compilation in \cite{Kraemer}:
the first group of 6 values is calculated in the LO collinear
approximation. In the second group effects of higher orders have been
taken into account approximately by using a parton shower Monte Carlo model
(PYTHIA). The third group uses a $k_t$ distribution for the gluon and
the last value in  Figure \ref{me} is derived in the $k_t$ factorisation 
approach.

The theoretical band in Figure \ref{fig4}a is calculated for
$(1<M_k<10)\cdot 10^{-2}\,\gev^3$ and 
$(0.3<{\cal O}^{\psi}\left[8,^3S_1\right]<2.0)\cdot10^{-2}\,\gev^3$ almost 
covering the full range of uncertainty.
The strong rise predicted at large $z$ values ($z\gtrsim 0.7$) is not seen in 
the data which has led to several theoretical attempts for an  explanation. 
In Figure \ref{fig4}b calculations of a different group~\cite{kniehl} are 
compared to the same data as in \ref{fig4}a.
The authors of~\cite{kniehl} attempt to estimate effects of higher orders. 
Here the data are a factor 3 above the predicted sum of singlet and octet contributions for
$z \lesssim 0.7$, while in Figure \ref{fig4}a the data are below the final
calculation. The rise in $z$ due to CO contributions in Figure \ref{fig4}b 
 takes place at $z \gtrsim 0.9$. This large $z$ region is dominated 
by diffraction and no experimental results on inelastic processes 
 are available. An intriguing 
question is of course a possible relation
of such CO contributions with what is traditionally attributed to
diffraction. 

\begin{figure}
\unitlength1.0cm
\begin{center}
\begin{picture}(24,10.5) 
\put(0.,0.){\epsfig{file=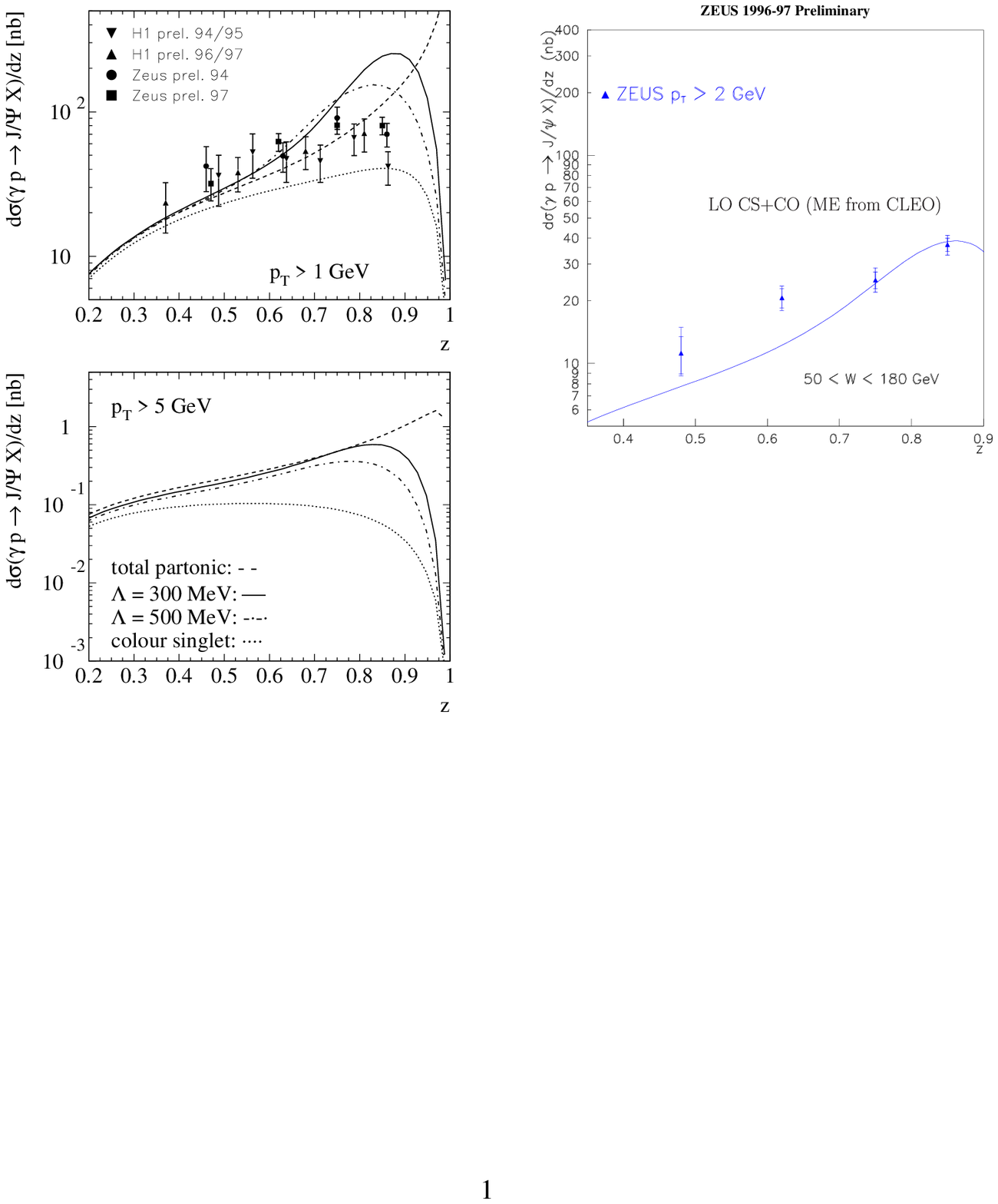,width=1.\textwidth}}
\end{picture}
\end{center}
\vspace{-0.5cm} 
\caption{\label{fig6} Left panels:
 Resummation of NRQCD contributions near $z \simeq 1$ according to \cite{wolf}.
In the upper panel predictions for \ptr$>1\gev$ are shown and compared to data. In the lower panel a cut of \ptr$>5\,\gev$ is applied. right panel: 
Comparison of the cross section \dsdz\ for \ptr$>2\,$\gev\ from the ZEUS Collaboration with a prediction from \cite{wolf} using shape functions and $\Lambda=300\,$MeV.}
\end{figure}

The NRQCD calculations shown in Figure \ref{fig4}a and b neglect the energy 
smearing of the transition of the $c\overline{c}$ state to the $J/\psi$ via
emission of soft gluons. This effect is expected to be large 
at $z\rightarrow 1$. 
In \cite{wolf}, an attempt was made to calculate
this smearing using the technique of shape functions known from calculations 
in decays of $B \rightarrow J/\psi+X$. The calculations for the $z$
distributions are shown together with the data in Figure  \ref{fig6} 
\footnote{The authors also 
extract the long range transition matrix elements fom B decays. The values, 
which were used for Figure \ref{fig6}, 
agree approximately with the values
derived from the $p_t$ distribution of $J/\psi$ mesons in $p\overline{p}$
collisions taking into account higher orders via a PYTHIA Monte
Carlo simulation \cite{Cano}.}.
In this comparison the standard cut $p_t >$ 1 GeV is applied. The smeared 
calculations are given for two values of $\Lambda$ and 
 show the expected decrease towards $z\ra 1$ while 
 the unsmeared curve (labelled ``total partonic'' in the figure) is seen to 
rise. However before decreasing the smeared curves are observed to increase 
above the unsmeared one (around $z\gtrsim 0.5$). 
The authors of \cite{wolf} conclude that the theory is not well behaved  
and suggest applying a higher $p_t$ cut.
Their prediction for $p_t >5\,\gev\ $is shown in the 
lower part of the figure. A comparison of this calculation with ZEUS results 
for a cut $p_t >$ 2 GeV is shown in the right panel of Figure \ref{fig6} 
and is seen to give an acceptable description of the data.

\begin{figure}[p]
\unitlength1.0cm
\begin{picture}(16,12.)
\put(-1.,0.5){\epsfig{file=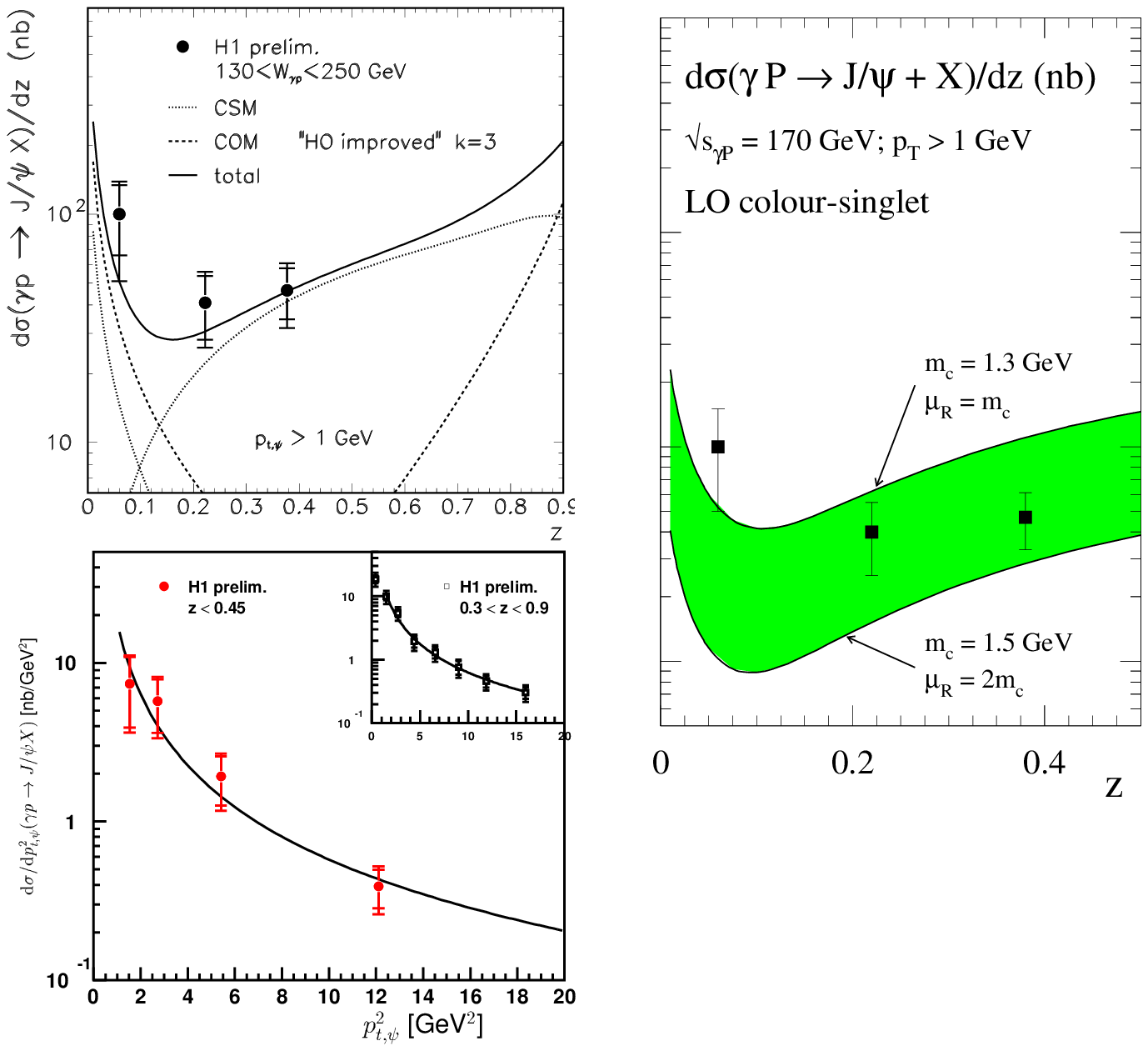,width=.8\textwidth}} 
\put(4.5,11.5){a)}
\put(10.5,9.5){b)}
\put(0.5,2.1){c)}
\end{picture}
\vspace{-1.5cm}
\caption{\label{lowz}
 Photoproduction of $J/\psi$ mesons at low $z$ from H1. Note the changed 
\wgp\ range compared to medium $z$. The theoretical predictions for the $z$ 
distribution in a) are from 
Kniehl \etal\cite{kniehl}. In b) the same H1 data are shown with a LO 
calculation within the CSM from \cite{Kraemer} depicting the major 
uncertainties. c) $d\sigma/d\ptt$\ in the low $z$ regime. The curve is a 
scaled fit to the medium $z$ data.}

\unitlength1.0cm
\begin{picture}(19,9)
\put(0,0.2){\epsfig{file=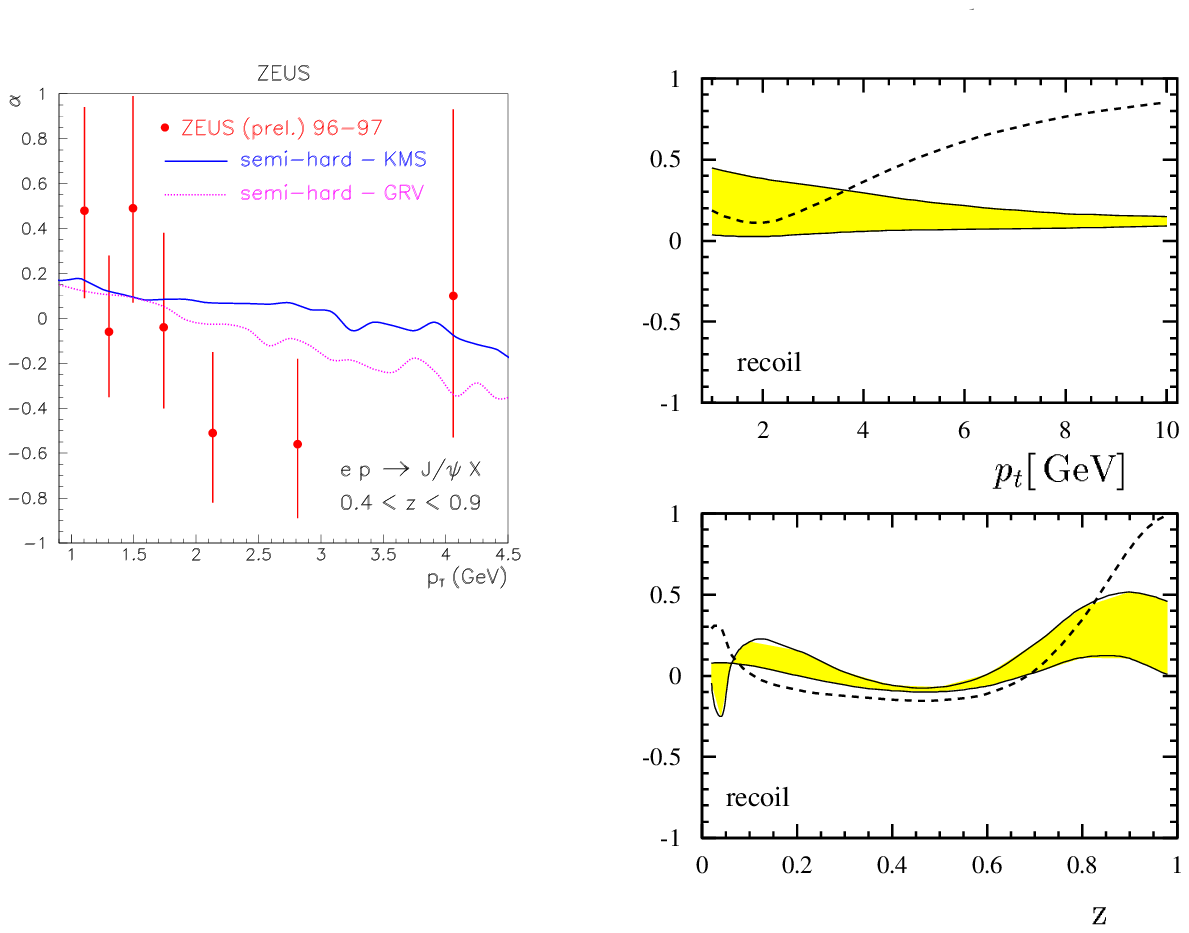,width=12cm}} 
\put(1.,4.){a)}
\put(11.,5.3){b)}
\put(6.,7.5){$\alpha$}
\end{picture}
\vspace*{-0.5cm}
\caption{\label{fig7}
a) The polarisation parameter $\alpha$ as a function of the transverse 
momentum of the  $J/\psi$ meson measured by the ZEUS 
collaboration~\cite{zeuspol}. 
b) Theoretical predictions for $\alpha$ within the CSM (dashed curve) and the 
Colour Octet Model as functions of $\ptr$ and $z$\cite{com}.
Note that in the calculation of the polarisation different reference 
axes have been used in a) and b), see text!}
\end{figure}

\subsection{Resolved Contributions and Polarisation Analysis}

The region of low $z$ values, $z<0.45$, is explored in an analysis of H1 
covering a region $130<\wgp<250\,\gev$ and $\ptr>1\,\gev$. The result
 is shown in Figure \ref{lowz}a in comparison to predictions
from \cite{kniehl}. At such low $z$ values, resolved photon contributions are 
expected and CO contributions are much larger than CS contributions.
Here the smearing effects of soft gluon emission are not expected to play
an important r\^{o}le. The number of parameters entering this comparison
is larger and probably a mere cross section measurement
will not suffice to distinguish between models. This can be judged from 
Figure \ref{lowz}b, where the predictions within the CSM are shown with the
dominant uncertainties, which overall amount to a factor 5.
The $p_t$ distribution of the data is shown in Figure \ref{lowz}c
and looks -- at least with present errors -- very similar to the
form found at medium $z$ values shown as a scaled curve.

Measurements of \jpsi\ polarisation in $p\overline{p} 
\rightarrow J/\psi + X$ were advocated to give independent proof of the 
presence of colour octet contributions. 
The polarisation is measured via the angular distribution of the \jpsi\ 
decay leptons in the rest frame of the $J/\psi$ meson. 
Parametrising it as $1+\alpha\,\cos^2\theta$,
$\alpha=+ 1(-1)$ corresponds to transverse (longitudinal) polarisation, 
while $\alpha= 0$ reflects no polarisation. 
The data for $J/\psi$ and $\psi'$ from the CDF collaboration~\cite{fail},
though limited in statistics, do not show the expected increase of the
 polarisation parameter $\alpha$ towards higher $p_t$ predicted within 
NRQCD\cite{Affolder:2000nn}.

ZEUS performed
a first measurement of the $J/\psi$ polarisation in $\gamma p$ scattering.
Using the flight direction of the $J/\psi$ in the
laboratory system as a reference axis the results in Figure \ref{fig7}a
were obtained. They are compared with the predictions of the ``semihard''
model \cite{baranov} using  unintegrated gluon density distributions 
based on BFKL evolution equations.
The trend of the data is described within this model.
There are no calculations within NRQCD/factorisation which one can compare 
directly with the data. In Figure \ref{fig7}b, calculated in the ``recoil'' 
system which uses the \jpsi\ direction in the $\gamma p$ system as a 
reference axis, one can however see that the 
$\alpha$ parameter increases with \ptpsi\ for the \csm\ while it remains 
small for the octet contributions. 

Summarising, there is still no evidence of colour octet contributions in 
photoproduction at HERA. Due to the experimental and theoretical uncertainties they can however not be excluded.

\section{Diffractive Production of $J/\psi$ Mesons}
\begin{figure}[p]
\unitlength1.0cm
\begin{picture}(19,8) 
\put(0,0){\epsfig{file=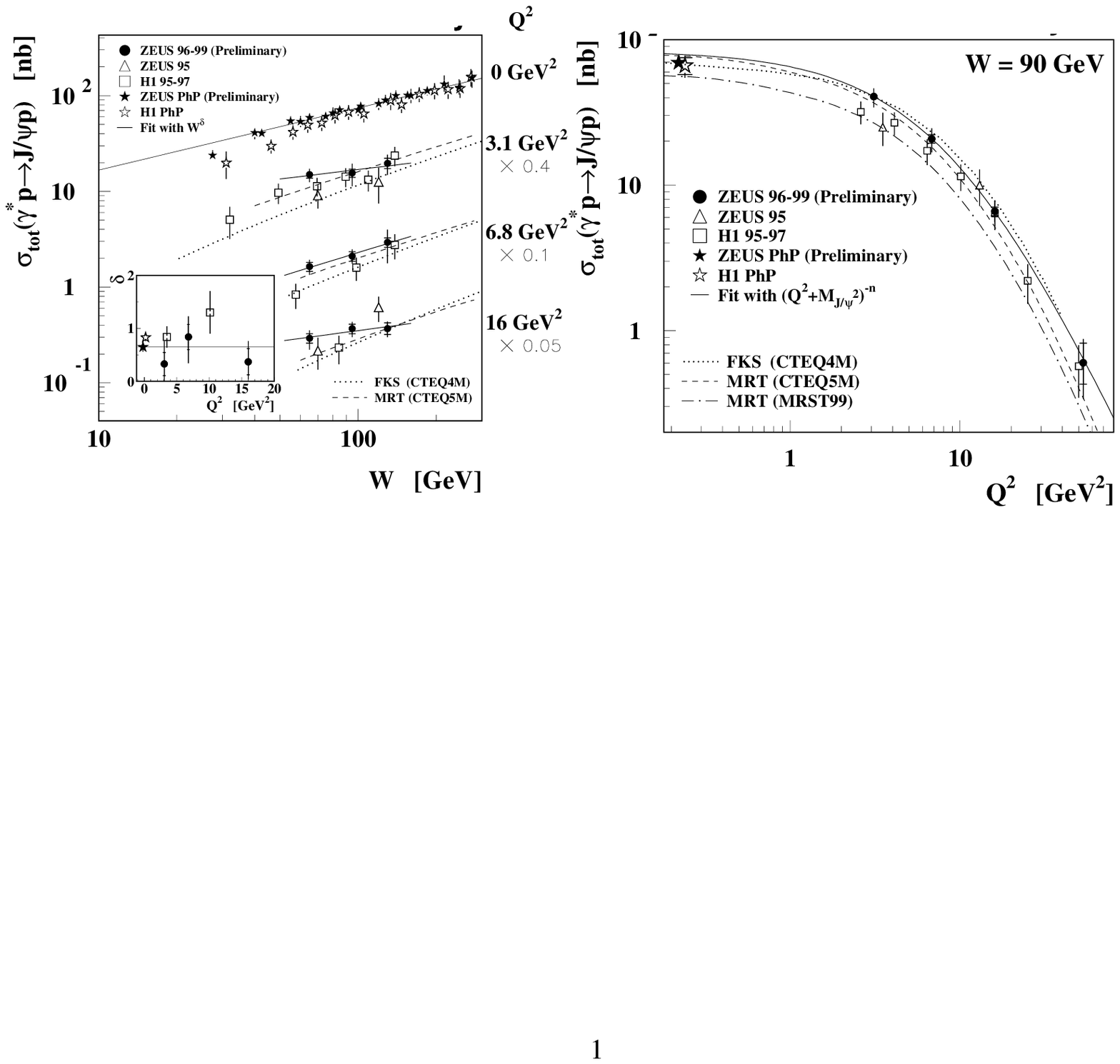,width=16cm,clip=}}
\end{picture}
\caption{\label{eladis} Elastic photoproduction of $J/\psi$ mesons as a function of \wgp\ (left) and \qsq\ (right). The data from H1 and ZEUS are shown with fits to the data as well as predictions within two  pQCD models.}

\begin{picture}(19,9.5) 
\put(0,-0.5){\epsfig{file=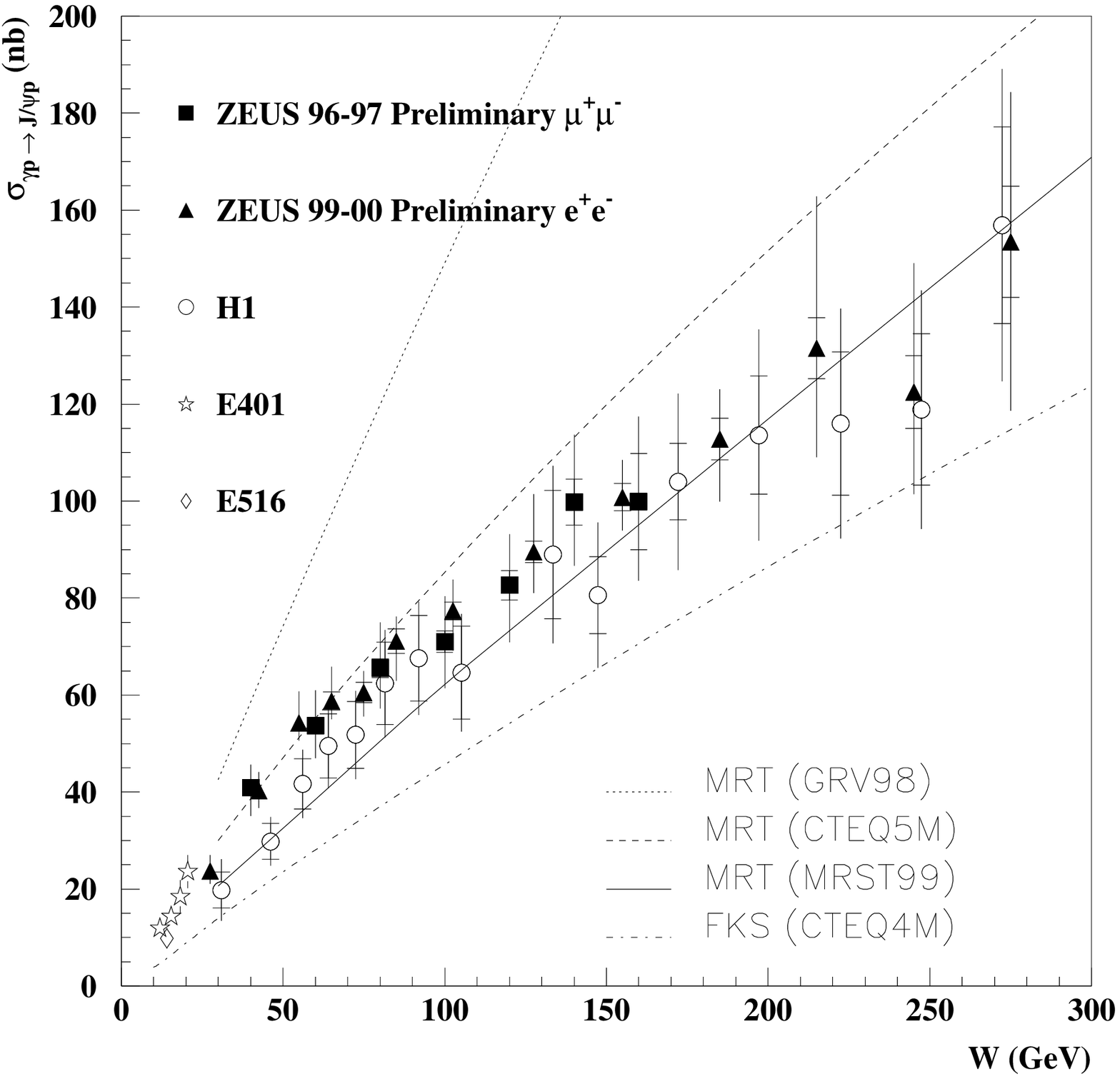,width=10cm,bbllx=0pt,bblly=0pt,bburx=530pt,bbury=530pt,clip=}}
\end{picture}
\caption{\label{elath}
Elastic photoproduction of $J/\psi$ mesons: the data from H1 and ZEUS are 
shown with 
two different theoretical predictions within pQCD using different parton 
distributions.}
\end{figure}

In the last few years it has been demonstrated that elastic $J/\psi$ production
can be described by perturbative QCD in photoproduction and also at finite 
$Q^2$. Data for elastic
diffractive processes are now available in a large kinematic range.
A summary of the published H1 data (21~$\picb$) 
\cite{Adloff:1999zs,Adloff:2000vm} and the new
preliminary data from ZEUS \cite{zeusel} (corresponding to $\sim$40~$\picb$) 
are shown in Figure \ref{eladis}. The data cover the regime of 
photoproduction, $Q^2 \rightarrow0$, extending almost to the kinematic
limit in the photon proton energy $W_{\gamma p} \leq 300\ \mbox{GeV}$.
 In $Q^2$ the data go up to $\sim 100\ \mbox{GeV}^2$.

The curves in Figure \ref{eladis} show fits to the data of the form 
$W^\delta$ and $\delta$ is also plotted as a function of $Q^2$ in the insert 
of Figure \ref{eladis}. The value of 
$\delta$ is on average well above $\sim 0.22$ expected from
Regge-type fits with a soft pomeron. The $Q^2$ distribution is fitted
by a functional form $(Q^2 + M^2_\psi)^{-n}$, yielding $n = 2.38 \pm 0.11$
for H1 and $n = 2.60 \pm 0.11^{+0.08}_{-0.05}$ for the ZEUS data.

The photoproduction data in Figure \ref{elath} are overlaid with predictions
from two theoretical groups using different approaches within 
pQCD~\cite{fks,mrt}. 
The basic picture is the exchange of two gluons between the proton and the 
$\ccbar$ pair. Apart from a number of technical differences the groups 
handle the conversion to a \jpsi\ meson differently: 
MRT use parton-hadron duality, while FKS use a wave function. 

The important prediction is for the slope of the data which is described 
well by all calculations shown with the exception of 
GRV98 partons. H1 has previously also found good agreement within the 
FKS calculations using CTEQ4M or MRSR2 partons~\cite{Adloff:2000vm}.

\begin{figure}
\unitlength1.0cm
\begin{picture}(19,9) 
\put(0,-0.5){\epsfig{file=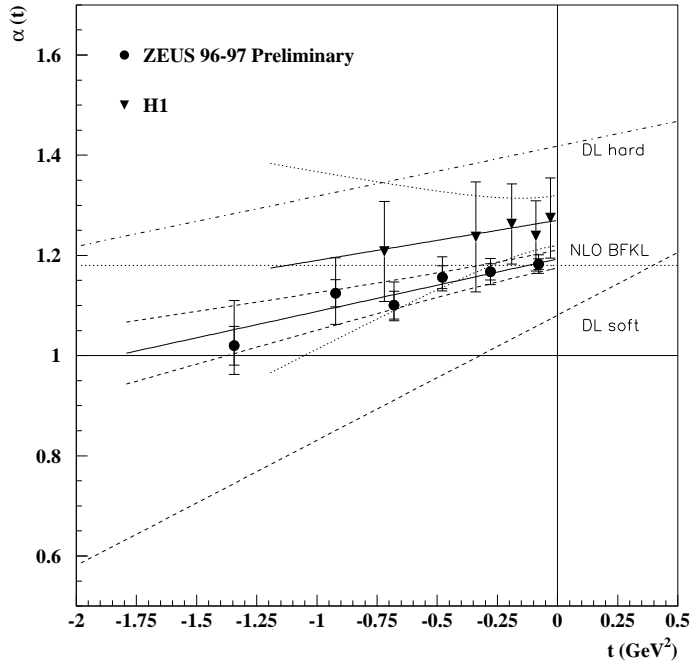,width=10cm,clip=}}
\end{picture}
\caption{\label{traj}
The effective trajectory for photoproduction of $J/\psi$ mesons extracted from photoproduction data. }
\end{figure}

Both collaborations have extracted the effective trajectory from 
photoproduction data. The procedure is to fit the data at fixed values
of $t$ with a form $W^{(\alpha (t) -1)}$. The resulting values of 
$\alpha(t)$ are displayed in
Figure \ref{traj} with separate fits to a simple form $\alpha(0) + \alpha't$.
The more accurate ZEUS data yield $\alpha(t) =
(1.193\pm0.011^{+0.015}_{-0.010})+(0.105\pm 0.024^{+0.022}_{-0.020})\cdot t$. 
The H1 data are compatible with this result.
The soft pomeron trajectory is ruled out. A NLO BFKL calculation  of the
intercept $\alpha (0) $\cite{brodsky} is in agreement with the data.

\begin{figure}
\unitlength1.0cm
\begin{picture}(20,8.9) 
\put(0,0){\epsfig{file=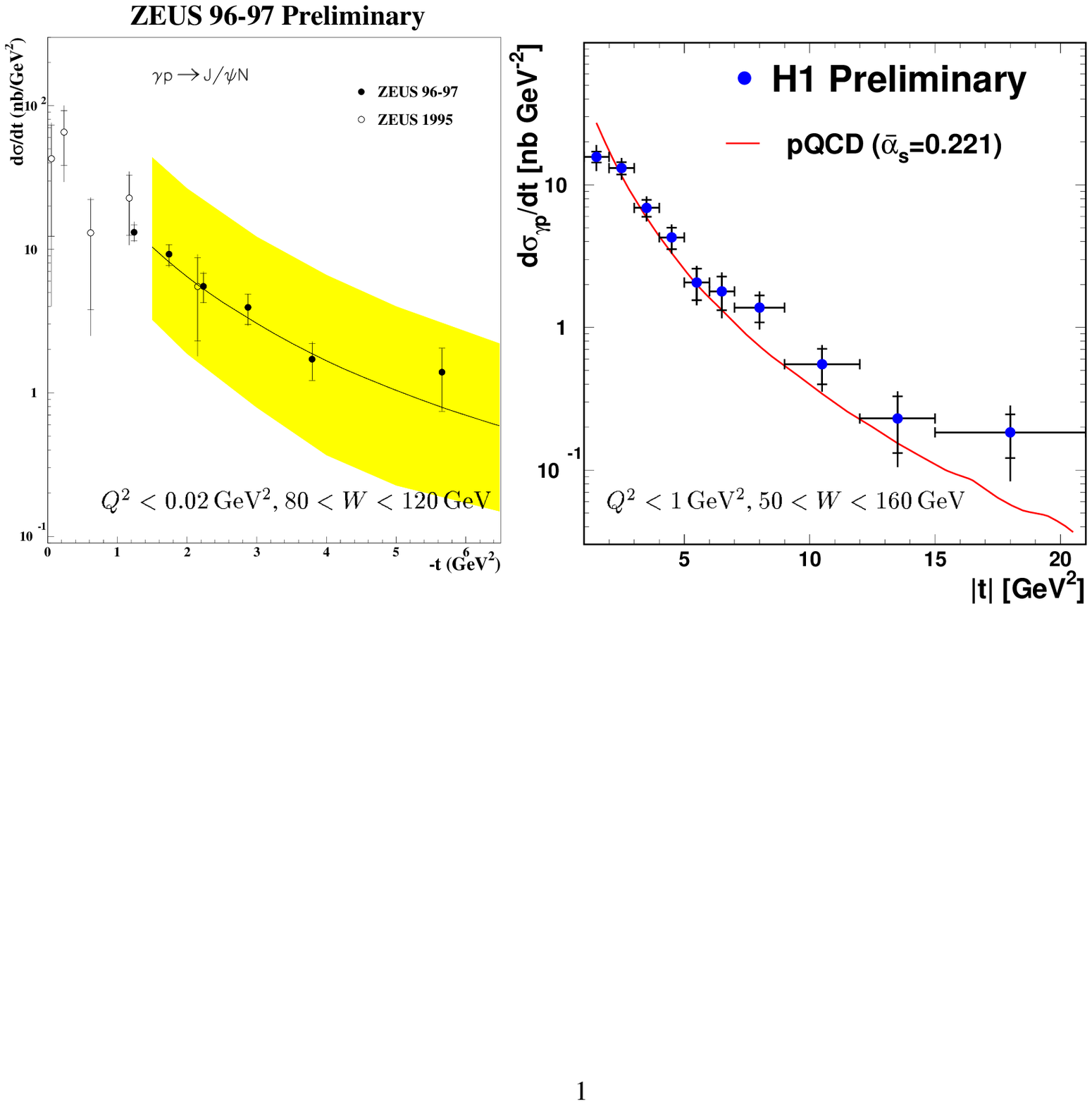,width=16cm,clip=}}
\end{picture}
\caption{\label{hight}
 $J/\psi$ Production at high values of $|t|$ in comparison with predictions from \cite{Bartels:1995rn} using BFKL dynamics. The band around the curve on the left depicts the major uncertainties. }
\end{figure}

The data shown so far are dominated by low values of $|t|$ since they
are found to drop with an exponential behaviour: $e^{-b|t|}$ with $b \approx
4\ \mbox{GeV}^{-2}$. Data collected at high $|t|$ are shown in 
Figure \ref{hight}.
At high values of $|t|$, the proton dissociates into a low mass system $X$.
ZEUS and H1, at present, measure slightly different cross sections: 
the ZEUS data
are corrected to a range $0.01<t/(t-M_X^2)<1$ while H1 corrects to the full 
kinematically allowed region of $M_X$. A LO calculation implementing the BFKL 
evolution equation  \cite{Bartels:1995rn} reproduces the data sets well using 
a value for $\overline{\alpha}_s \approx 0.2$. 

\section{Summary and Outlook}
Inelastic photoproduction of \jpsi\ mesons at HERA continues
to be well described by the COlour Singlet Model calculated in 
next-to-leading order.
At a \ptpsi$\simeq 4\,\gev$, the measured cross section is reproduced by 
the NLO calculation, which is a factor of 10 above the LO prediction.
A reasonable alternative description of the data on a Monte Carlo basis can 
be obtained using parton distributions not integrated over the 
transverse momentum based on the CCFM evolution equations.
As regards Colour Octet contributions, only LO calculations are available.
Due to normalisation uncertainties, the comparison with data at high $z$ 
values is still ambiguous. NRQCD resummations at high $z$ seem to 
necessitate a higher \ptpsi\ cut and then describe the data reasonably well.
A first analysis of data at low $z$ values shows sensitivity to resolved 
contributions. However since in this regime more parameters enter, a mere 
cross section measurement may not be sufficient to decide on Colour Octet 
contributions.
A first attempt to measure the \jpsi\ meson polarisation looks promising.
This and all other analyses will profit from the increased statistics which will soon become available.  

For elastic \jpsi\ production,
the \wgp\ range in photoproduction has been extended almost to the 
kinematic limit. The cross section continues to rise. Parametrising it  
as $\wgp^\delta$, $\delta$ is measured to be large, of order 0.8. 
The data at higher \qsq\ are in rough agreement with such a steep rise. 
Comparisons of theoretical calculations within pQCD show a sensitivity to 
the gluon distribution.
A ``trajectory'' measurement excludes the soft pomeron trajectory.
The measurement of diffractive processes has been extended to $|t|$ values of 
$\sim 20\,\gevt$. The data is well described by calculations based on the BFKL equation.

\paragraph*{Acknowledgement}
I wish to thank the organizers for letting me participate in this very 
fruitful and stimulating meeting in beautiful surroundings. Thanks to all 
`theoretical' and `experimental' colleagues for discussions about their 
data or theories!\\

\end{document}